\documentclass[aps,twocolumn,showpacs]{revtex4}
\usepackage{graphicx}
\usepackage{dcolumn}
\usepackage{bm}

\begin{document}

\title{Stickiness in a bouncer model: A slowing mechanism for Fermi
acceleration}

\author{Andr\'e L.\ P.\ Livorati$^1$, Tiago Kroetz$^{2}$, Carl P.
Dettmann$^{3,4}$, Iber\^e Luiz Caldas$^{1}$ and Edson D.\ Leonel$^{4,5}$}
 
\affiliation{$^1$Instituto de F\'isica - IFUSP - Universidade de S\~ao
Paulo - USP  Rua do Mat\~ao, Tr.R 187 - Cidade Universit\'aria --
05314-970 -- S\~ao Paulo -- SP -- Brazil -- livorati@usp.br\\
$^2$ Departamento de F\'isica - Universidade Tecnol\'ogica Federal do
Paran\'a - UTFPR Campus Pato Branco - 85503-390 - Pato Branco - PR - Brazil
\\ 
$^3$ School of Mathematics - University of Bristol - Bristol BS8 1TW,
United Kingdom\\
$^4$Departamento de Estat\'istica Matem\'atica Aplicada e Computa\c c\~ao
-- UNESP -- Univ Estadual Paulista -- Av. 24A, 1515 - Bela Vista -
13506-900 - Rio Claro - SP - Brazil\\
$^5$Abdus Salam - ICTP, 34151 Trieste, Italy}


\begin{abstract}
Some phase space transport properties for a conservative bouncer 
model are studied. The dynamics of the model is described by using a
two-dimensional measure preserving mapping for the variables velocity and
time. The system is characterized by a control parameter $\epsilon$ and
experiences a transition from integrable ($\epsilon=0$) to non integrable
($\epsilon\ne 0$). For small values of $\epsilon$, the phase space shows a
mixed structure where periodic islands, chaotic seas and invariant tori
coexist. As the parameter $\epsilon$ increases and reaches a critical
value $\epsilon_c$ all invariant tori are destroyed and the chaotic sea
spreads over the phase space leading the particle to diffuse in velocity
and experience Fermi acceleration (unlimited energy growth). During the
dynamics the particle can be temporarily trapped near periodic and stable
regions. We use the finite time Lyapunov exponent to visualize this effect. 
The survival probability was used to obtain 
some of the transport properties in the phase space. For large $\epsilon$, the survival
probability decays exponentially when it turns into a slower decay as the
control parameter $\epsilon$ is reduced. The slower decay is related to trapping dynamics, 
slowing the Fermi Acceleration, i.e., unbounded growth of the velocity
\end{abstract}

\maketitle

\section{Introduction}
\label{sec1}

It is known that Hamiltonian systems are typical non-ergodic and
non-integrable \cite{ref1}. The phase space of such systems is divided into
regions with regular and chaotic dynamics. These dynamical regions are
connected by a layer, where regular or irregular motion, can or cannot mix,
depending upon on the number of degrees of freedom of the system, as well
properties of the limiting surface itself. Such a division leads to the
stickiness phenomenon \cite{ref2,ref3} which is manifested through the fact
that a phase trajectory in a chaotic region passing near enough a
Kolmogorov-Arnold-Moser (KAM) island, evolves there almost regularly during
a time that may be very long. However, when an orbit resides in a chaotic
region far from the set of KAM regions, it moves chaotically in the sense
that two nearby initial conditions apart from each other exponentially as
the time evolves. Therefore the stickiness of phase trajectories has a
crucial influence on the transport properties of Hamiltonian systems, and
its relation to physical systems is one of the most important open problems
of nonlinear dynamics \cite{ref4,ref4a}. Applications of stickiness can be
found in astronomy \cite{ref5}, fluid mechanics \cite{ref6}, Levy flights
\cite{ref7}, also in biology \cite{ref8}, in plasma physics
\cite{ref9,ref10} and many others.

One of the main consequences of the influence of orbits in sticky regime is
observed in the transport properties along the phase space. Therefore it
may give rise to the following question: May sticky orbits influence the
Fermi acceleration phenomenon? Fermi acceleration (FA) was
introduced by the first time in 1949 by Enrico Fermi \cite{ref11} as an
attempt to explain the possible origin of the high energies of the cosmic
rays. Fermi claimed that the charged cosmic particles could acquire energy
from the moving magnetic fields present in the cosmos. His original idea
generated a prototype model which exhibits unlimited energy growth and is
called the bouncer model. The model consists of a free particle (making
allusions to the cosmic particles) which is falling under
influence of a constant gravitational field $g$ (a mechanism to inject the
particle back to the collision zone) and suffering collisions with a heavy
and time-periodic moving wall (denoting the magnetic fields). The model
is characterized by a control parameter $\epsilon$ and has a transition from
integrability $\epsilon=0$ to non integrability $\epsilon\ne 0$. A mixed
structure of the phase space is observed for lower values of $\epsilon$ and
strong chaotic properties are present in the regime of large values of the
parameter, say $\epsilon>\epsilon_c$ where at $\epsilon_c$ the system
experiences a transition from local to globally chaotic regime
(destruction of invariant spanning curves).

In this paper we revisit the bouncer model seeking to understand and
describe some transport properties along the phase space particularly
focusing on the dynamics of sticky orbits. The model is described by a two
dimensional, nonlinear and measure preserving mapping for the variables
velocity of the particle and time at the collision with the moving wall. As
the parameter $\epsilon$ is increased, the number of islands in the phase
space decreases. For the regime of high nonlinearity $\epsilon\gg 1$, almost
no islands are observed. The temporarily trapping dynamics due to the
sticky regions are more often observed in the regime of small $\epsilon$
where a mixed structure of the phase space is present. We use the finite
time Lyapunov exponent spectrum of the orbits and a statistical analysis of
escape rates to investigate the influence of the stickiness in dynamics of
an ensemble of non interacting particles. We therefore conclude that the stickiness present in 
the system acts as a slowing mechanism for FA.

The paper is organized as follows: In Sec. \ref{sec2} the mapping that
describes the dynamics of the model is obtained. In Sec.\ref{sec3}, the
numerical results are present which include the calculation of the finite
time Lyapunov exponent and escape rates for the velocity as a function of
$\epsilon$. Finally, the conclusions and final remarks are drawn in
Sec.\ref{sec4}.

\section{The model, the mapping and chaotic properties}
\label{sec2}

We discuss in this section the procedures used to construct the mapping
that describes the dynamics of the system. The model consists of a
classical particle of mass $m$ which is moving in the vertical direction
under the influence of a constant gravitational field $g$. It also suffers
elastic collisions with a periodically moving wall whose position is
given by $y(t)=\varepsilon \cos(wt)$, where $w$ is the frequency and
$\varepsilon$ is the amplitude of oscillation respectively.

The dynamics of the system is made by the use of a two dimensional,
nonlinear and measure preserving mapping for the variables velocity of the
particle $v$ and time $t$ immediately after a $n^{th}$ collision of the
particle with the moving wall. During the dynamics, two distinct kinds of
collisions may be observed: (i) multiple collisions of the particle with
the moving wall -- those happening before the particle leaves the collision
zone (the collision zone is defined as the region
$y\in[-\varepsilon,\varepsilon]$) -- or; (ii) a single collision of the
particle with the moving wall (causing the particle to leave the collision
zone). Before writing the equations of the mapping, it is important to
mention there are an excessive number of control parameters, 3 in total,
namely $\varepsilon$, $g$ and $w$. We may define dimensionless and more
convenient variables as: $V_n=v_n w/g$, $\epsilon=\varepsilon w^2/g$ and
measure the time in terms of the number of oscillations of the moving wall
$\phi_n=w t_n$.

We assume that at the instant $\phi\in[0,2\pi]$ the position of the
particle is $y_p(\phi_n)=\epsilon\cos(\phi_n)$ with initial velocity
$V_n>0$, which lead us to obtain the following expression for the mapping
\begin{equation}
T_c:\left\{\begin{array}{ll}
V_{n+1}=-{V_n^*}+{\phi_c}-2\epsilon\sin(\phi_{n+1})\\
\phi_{n+1}=[\phi_n+\Delta T_n]~~{\rm mod (2\pi)}\\
\end{array}
\right.,
\label{eq1}
\end{equation}
where the index $c$ stands for the complete version of the model (the one
which takes into account the movement of the moving wall) and the
expressions for $V_n^*$ and $\Delta T_n$ depend on what kind of collision
happens. For case (i), i.e. the multiple collisions, the expressions are
$V_n^*=V_n$ and $\Delta T_n=\phi_c$ where $\phi_c$ is obtained from the
condition that matches the same position for the particle and the moving
wall. It leads to the following transcendental equation that must be solved
numerically
\begin{equation}
G(\phi_c)=\epsilon\cos(\phi_n+\phi_c)-\epsilon\cos(\phi_n)-V_n\phi_c+{{
1}\over{2}}\phi_c^2~.
\label{eq2}
\end{equation}

If the particle leaves the collision zone case (ii) applies. The expressions
are $V_n^*=-\sqrt{V_n^2+2\epsilon(\cos(\phi_n)-1)}$ and $\Delta
T_n=\phi_u+\phi_d+\phi_c$ with $\phi_u=V_n$ denoting the time spent by the
particle in the upward direction up to reaching the null velocity,
$\phi_d=\sqrt{V_n^2+2\epsilon(\cos(\phi_n)-1)}$ corresponds to the time that
the particle spends from the place where it had zero velocity up to the
entrance of the collision zone at $\epsilon$. Finally the term $\phi_c$ has
to be obtained numerically from the equation $F(\phi_c)=0$ where
\begin{equation}
F(\phi_c)=\epsilon\cos(\phi_n+\phi_u+\phi_d+\phi_c)-\epsilon-V_n^*
\phi_c+{{1}\over{2}}\phi_c^2~.
\label{eq3}
\end{equation}

\begin{figure*}[t]
\begin{center}
\centerline{\includegraphics[width=17cm,height=12.0cm]{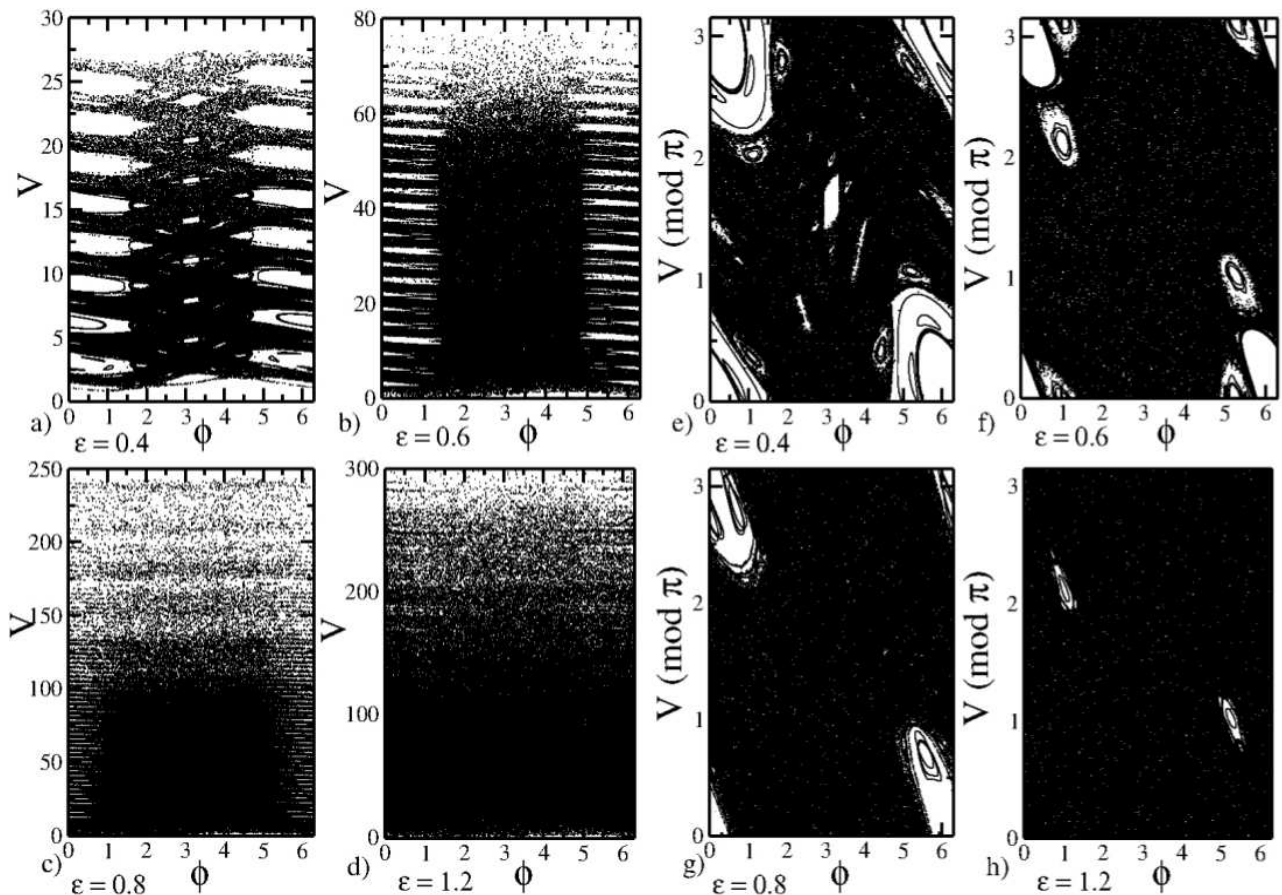}}
\end{center}
\caption{Plot of the phase space for the bouncer model considering the
control parameters: (a) and (e) $\epsilon=0.40$; (b) and (f)
$\epsilon=0.60$; (c) and (g)$\epsilon=0.8$; (d) and h) $\epsilon=1.20$.}
\label{fig2}
\end{figure*}

The extended phase space for the whole version of the model considers four
variables namely: (1) $x_w$ denoting the position of the moving wall; (2)
$V_p$ corresponding to the velocity of the particle; (3) $E_p$ which is the
energy of the particle and (4) the time $t$. The canonical pairs however
are: position and velocity $(x_w,V_p)$ and; energy and time $(E_p,t)$. As
the way the mapping was constructed, the variables used are not canonical
ones therefore the determinant of the Jacobian matrix is 
\begin{equation}
{\rm Det~J}=\left[{{V_n+\epsilon\sin(\phi_n)}\over{V_{n+1}
+\epsilon\sin(\phi_{n+1})}}\right]~,
\label{eq4}
\end{equation}
which is clearly different from unity as it should be if the canonical
pair was considered. However we may say that it preserves the following
measure in the phase space $d\mu=[V+\epsilon\sin(\phi)]dVd\phi$.

A common version which is also present in the literature is the so called
simplified version. It was proposed many years ago \cite{ref17} as an
attempt to keep the essence of the problem but at the same time allow
numerical computations to be realized in a reasonable time when computers
were far slower. Also it could reduce the complexity of the equations at a
level that analytical calculations could be obtained. It assumes that the
wall is fixed -- so that the calculation of the time between collision does
not evolve numerical solution of transcendental equations --, but at the
instant of the collision, the particle suffers an exchange of energy and
momentum as if the wall were moving. In this version, the extended phase
does not consider more the position of the moving wall, because by
definition it is fixed, causing the canonical pair to be the velocity and
time. The mapping is then written as
\begin{equation}
T_s:\left\{\begin{array}{ll}
V_{n+1}= | V_n-2\epsilon\sin(\phi_{n+1})|\\
\phi_{n+1}=\phi_n+2V_n~~{\rm mod (2\pi)}\\
\end{array}
\right.,
\label{eq5}
\end{equation}
where the modulus function is introduced to avoid the particle to move
beyond the wall. After a collision, if the particle has a negative
velocity, we re-inject it back with the same velocity. For the simplified
version and given the variables describing the dynamics are the canonical
pair, the determinant of the Jacobian is given by ${\rm Det~J}=\pm 1$. The
simplified version of the model also allow us to make a connection with the
so called standard mapping. Defining $I_n=2V_n$, $K=4\epsilon$ and 
$\theta_n=\phi_{n+1}+\pi$ the simplified version is written as the standard
mapping. The variation of the control parameter $\epsilon$ leads the
dynamics to experience a transition from locally to globally chaotic
dynamics
as similarly observed in the standard mapping \cite{ref33}. Indeed for
$\epsilon<\epsilon_c\approx0.2429$ the phase space has invariant spanning
curves (also called invariant tori) and unlimited energy growth, which
characterizes FA, is not observed. As the parameter $\epsilon$ is increased,
the fixed points become unstable and bifurcate for $\epsilon>1$ ($K>4$).

The period-1 fixed points are obtained solving the two equations
simultaneously $V_{n+1}=V_n=V^*$ and $\phi_{n+1}=\phi_n=\phi^*$ and are
given by
\begin{eqnarray}
V^*&=&\pi l~~~,~~~ l=0,1,2,\ldots\label{eq7a}\\
\phi^*&=&\arcsin \left({\pi m \over 2\epsilon} \right)~~~,~~~m=0,1,2,\ldots
\label{eq7}
\end{eqnarray}
Thus, there are windows of periodicity for the period one fixed points which
depend on $\epsilon$. The linear stability for these fixed points are given
by
\begin{equation}
(2\pi)^2(p-1)^2<16\epsilon^2<(2\pi)^2(p-1)^2+4~,
\label{eq8}
\end{equation}
where $p=l-m$.

Figure \ref{fig2} shows the structure of the phase space for the complete
version of the bouncer as a function of the control parameter $\epsilon$.
The accuracy used to solve numerically both $F$ and $G$ was $10^{-12}$
using the bisection method. As $\epsilon$ is increased the stable regions
(mainly marked by periodic fixed points) reduce leading the phase space to
have large unstable regions. The regions of sticky are more often observed
for smaller values of $\epsilon$ due to the existence of many islands in the
phase space as compared to large values of $\epsilon$. Analyzing Fig.
\ref{fig2} we see that the phase space has a repeating structure in $\pi$
in the velocity axis. Thus, let us plot the phase space taking the
$mod(\pi)$
for velocity. Such a plot is useful for observing the
evolution of the fixed points and the possible trappings caused by sticky
orbits. The control parameters used to construct Fig. \ref{fig2} were: (a)
and (e) $\epsilon=0.40$; (b) and (f) $\epsilon=0.60$; (c) and (g)
$\epsilon=0.80$; and (d) and (h) $\epsilon=1.20$. For each figure a set of
$100$ different initial conditions were evolved in time until $10^5$
collisions with the moving wall. The initial velocity was chosen such
that its minimum value was higher than the stable region in
$V\in[0,2\epsilon]$.

\begin{figure}[t]
\includegraphics[width=1.0\linewidth]{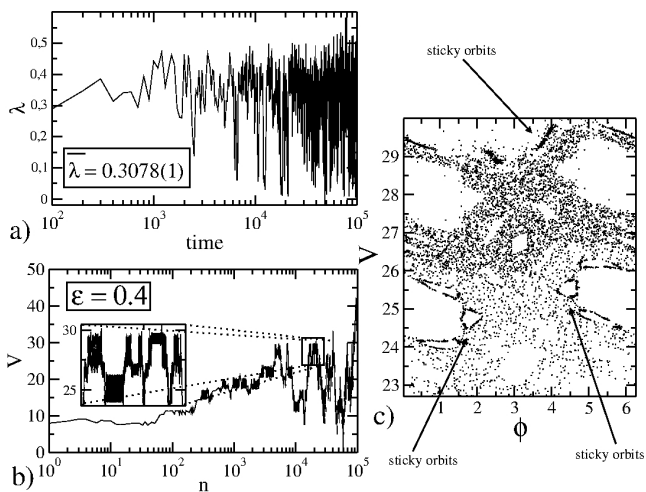}
\caption{(a) Plot of the FTLE for an initial condition chosen in the
chaotic sea. (b) shows the evolution of the same initial condition of (a)
for a plot of velocity against the number of collisions. (c) the zoom-in
window of the previously selected area of (b) showing these trapping orbits
in the phase space coordinates $(V,\phi)$.} 
\label{fig3}
\end{figure}

\section{Numerical Results}
\label{sec3}

This section is divided in two parts. In the first one we discuss the
results for the Lyapunov exponent obtained at finite time while in the
second we present our discussions and show results for orbits that survive
longer the dynamics after being trapped by some sticky regions.

\subsection{Lyapunov exponents}

Let us start discussing our results for the positive Lyapunov exponent for
chaotic components of the phase space. The Lyapunov exponent has been widely
used to quantify the average expansion or contraction rate for a small
volume of initial conditions. If the Lyapunov exponent is positive, the
orbit is said to be chaotic leading to an exponential separation of two
nearby initial conditions. On the other hand, a non positive Lyapunov
exponent indicates regularity and the dynamics can be in principle
periodic or quasi-periodic. The Lyapunov exponents are defined as follows
\cite{ref35} (see for example \cite{ref36} for applications in higher
dimensional systems):
\begin{equation}
\lambda_j = \lim_{n\rightarrow\infty} {1\over{n}} \ln |{\Lambda_j^n}|,
~~~j=1,2~,
\label{eq12}
\end{equation}
where  $\Lambda_j^n$, are the eigenvalues of the matrix $M={\prod_{i=1}^n}
J(V_i,\phi_i)$ and $J_i$ is the Jacobian matrix evaluated over the orbit.

\begin{figure}[t]
\includegraphics[width=1.0\linewidth]{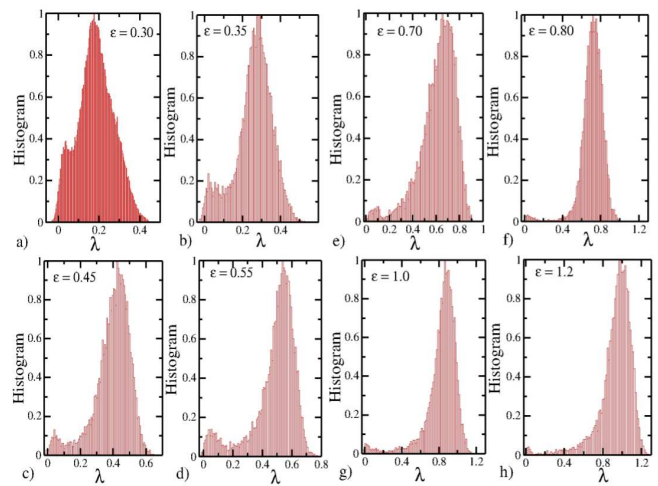}
\caption{(Color online). Plot of the FTLE distributions for several values
of the parameter $\epsilon$. One sees two distinct peaks, a larger one
representing the mean value of the Lyapunov exponent, and the secondary one,
is due to orbits in stickiness regime. As $\epsilon$ increases, the
magnitude of the secondary peak decreases indicating that for higher
values of $\epsilon$, less sticky orbits are observed. The control
parameters used were: (a) $\epsilon=0.3$; (b) $\epsilon=0.35$; (c)
$\epsilon=0.45$; (d) $\epsilon=0.55$; (e) $\epsilon=0.70$; (f)
$\epsilon=0.80$; (g) $\epsilon=1.0$; (h) $\epsilon=1.2$.} 
\label{fig4}
\end{figure}

\begin{figure*}[t]
\begin{center}
\centerline{\includegraphics[width=16cm,height=12.0cm]{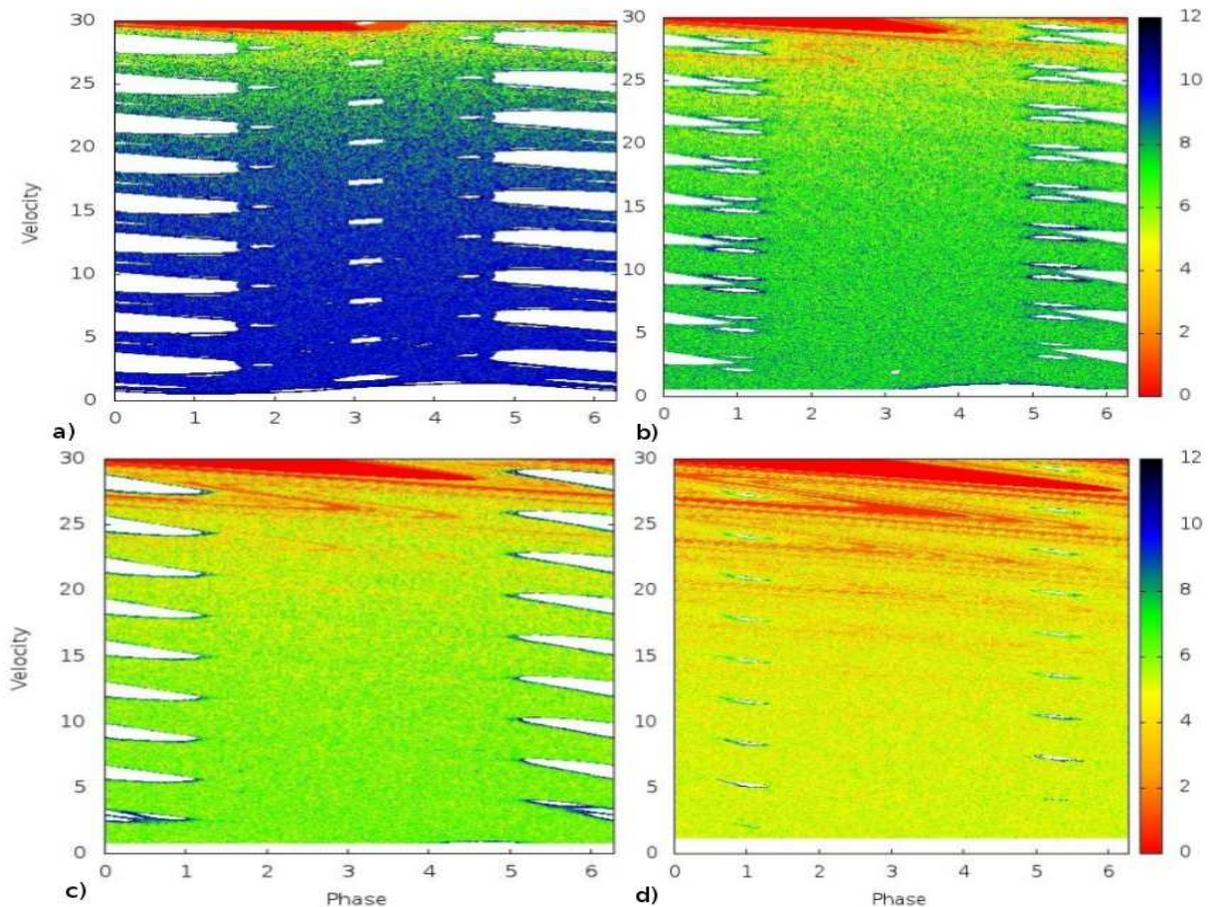}}
\end{center}
\caption{(Color online) Plot of the time evolution of initial conditions to
reach a hole at $V_{\rm hole}=30$. The control parameters used were:
(a)$\epsilon=0.4$, (b)$\epsilon=0.6$, (c)$\epsilon=0.8$ and (d)
$\epsilon=1.2$. Dark blue indicates long time evolution until reaching the
hole while red indicates fast scape. White denotes the particle never
escaping until $10^5$ collisions.}
\label{fig5}
\end{figure*}

\begin{figure*}[t]
\begin{center}
\centerline{\includegraphics[width=17cm,height=12.0cm]{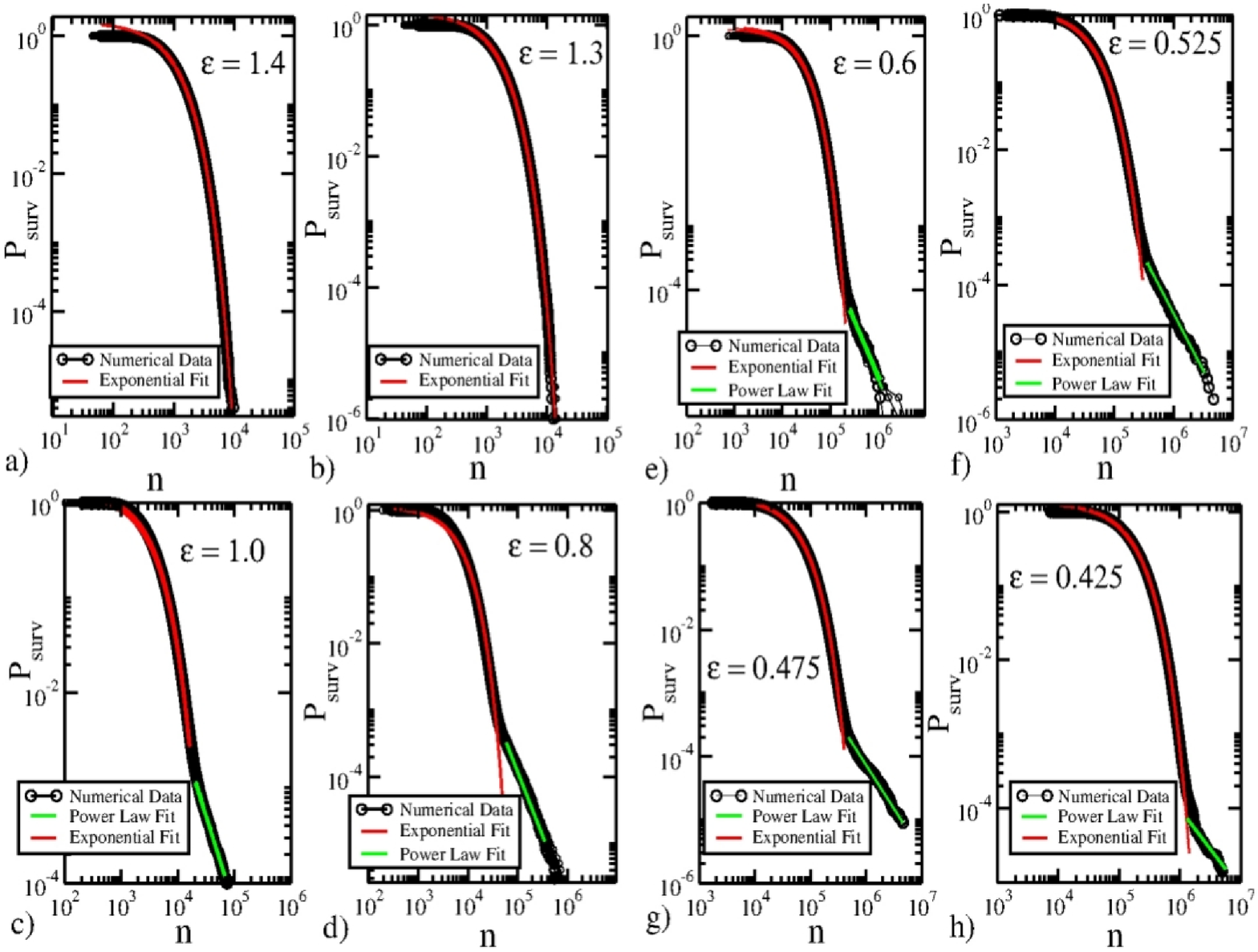}}
\end{center}
\caption{(Color online). Plot of the curves of $P_{surv}$ for different
control parameters. One sees the change of the behavior of $P_{surv}$ as the
parameter $\epsilon$ is decreased. The control parameters used were:
a)$\epsilon=1.4$, b)$\epsilon=1.3$, c)$\epsilon=1.0$, d) $\epsilon=0.8$,
e)$\epsilon=0.6$, f)$\epsilon=0.525$, g)$\epsilon=0.475$ and 
h)$\epsilon=0.425$.}
\label{fig6}
\end{figure*}

In the dynamics of the bouncer model, chaotic and regular motion can
coexist in the phase space, which introduces large variations and local 
instability along a reference chaotic trajectory. Such variations, are
related to alternations between different motions, in a qualitative way of
saying, as well as chaotic and quasi-regular motions. In order to
characterize such peculiar variation dynamics, we used the Finite-Time
Lyapunov Exponent (FTLE) \cite{ref27}. Once the trappings caused by orbits
in stickiness regime happen just for a finite time, this technique is useful
to quantify the trapping effects. It was shown \cite{ref27} that when the
FTLE distributions present small values, it is related to existence of
long-lived jets from a two-dimensional model for fluid mixing and transport.
This can be understood, in a dynamic point of view as stickiness
trajectories in the phase space.

Figure \ref{fig3}(a) shows the evolution of the FTLE, for an initial
condition chosen in the chaotic sea, for $\epsilon=0.4$. One sees a very
irregular behavior along the time, alternating average contractions and
repulsions, leading to and average value as $\bar{\lambda}=0.3078(1)$. In
Fig. \ref{fig3}(b) it is shown the evolution of the same initial condition
of Fig. \ref{fig3}(a) however plotted the velocity as a function of the
number of collisions. It is clear in Fig. \ref{fig3}(b) the successive
trappings along the orbit, and how they ``slow down'' the energy growth,
that characterizes the FA. Also, we set a zoom-in window in
Fig. \ref{fig3}(b) and plot the corresponding orbit in the phase space
portrait $(V,\phi)$, in order to identify some of these stickiness orbits
in Fig. \ref{fig3}(c).

To optimize the window of time to be used in the FTLE calculations, we have
considered different lengths in several simulations. After some
comparisons of the results we come up, based in fluctuations of the Lyapunov
exponents, to a finite time of $100$ collisions that was then used to study
the distribution of FTLE. It is known in the literature \cite{ref27}that the FTLE
distribution has a Gaussian shape, where the large peak can be interpreted
as the mean value of the Lyapunov exponent. If the system presents any
periodic or quasi-periodic motion, besides  chaos in its dynamics, the FTLE
distribution can have a secondary peak in the region of very low value of
the Lyapunov exponent. Such secondary peak is interpreted as sticky orbits
along the dynamics evolution \cite{ref36,ref27} responsible for trapping
the dynamics. The distribution for several FTLE are shown in Fig. \ref{fig4}
for different control parameter $\epsilon$ as labeled in the figure. We can
see from Fig. \ref{fig4} that the secondary peak of the FTLE distribution
is more evident for small values of $\epsilon$. Just to have a glance of
the influence of the second peak in the distribution represents up about 
$20\%$ of the whole distribution of Fig. \ref{fig4}(b). The fraction of the
distribution of the FTLE for the secondary peak decreases as $\epsilon$ is
increased. Such a result is expected because for higher values of
$\epsilon$ less islands in the phase space are observed as previously shown
in Fig. \ref{fig2}.

\subsection{Survival Probability and Escape Rates}

In this section we discuss results for orbits that survive
until reaching a pre-defined velocity at which they are assumed
to escape. To do that we consider the existence of a hole in the velocity 
coordinate of the phase space. If the particle reaches such a velocity
or higher, its dynamics is stopped and a new initial condition is started.
The introduction of the hole allow us to study transport properties as well
as characterize, through statistical analysis of survival probability and
time-correlation decays, the influence of sticky orbits along the dynamics
of the model \cite{ref37,ref38,ref39}.

To study the transport properties, we set a grid of initial conditions
equally distributed along the velocity and phase. Indeed a grid of
$500\times500$ initial conditions with $V_0\in[\epsilon,30]$ and
$\phi_0\in[0,2\pi]$ were considered. Then each initial condition was evolved
in time up to the limit of $10^5$ collisions with the moving wall or until a
hole placed in the velocity axis at $V_{\rm hole}=30$ is reached. Figure
\ref{fig5} shows a plot of the initial conditions evolved until $10^5$
collisions with the moving wall or up to the particle reaching the hole.
The color ranging from red (fast escape) to blue (long time dynamics)
denotes
the time (plotted in logarithmic scale) the particle spends until reaching
the escape velocity. White regions denote that the particle never escaped. The
control parameters used to construct the figures were: (a) $\epsilon=0.4$;
(b) $\epsilon=0.6$; (c)$\epsilon=0.8$ and; (d) $\epsilon=1.2$.

We see from Fig. \ref{fig5}(a) where $\epsilon=0.40$, that low initial
velocities spend large time accumulating energy until reach the hole at
$V=30$. Additionally one sees many stability islands where the orbits can
get temporally trapped and been released after a while. These temporally
trappings are caused by sticky regions. Such dynamical regimes can be
visualized by the dark regions marked by blue color in Figs.
\ref{fig5}(b,c) whose control parameters are respectively $\epsilon=0.6$ and
$\epsilon=0.8$. When the control parameter $\epsilon$ is raised, the
particles reach the hole faster as we can see from Figs. \ref{fig5}(b,c,d).
In particular for Fig. \ref{fig5}(c) one sees that the first stability
island disappeared. The stability regions are getting smaller and smaller as
the control parameter $\epsilon$ raises and from Fig. \ref{fig5}(d) they
appear to be very small for $\epsilon=1.2$. However even for a control
parameter where the stability islands are small, we see that the sticky
orbits are still present and indeed are marked by the dark blue color in the
plot.

The statistics of the cumulative recurrence time distribution which is
obtained from the integration of the frequency histogram distribution for
the escape can also be obtained. To do that we consider now that the escaping
velocity is set as $V_{\rm hole}=100$ although any other velocity could be
considered. Their cumulative recurrence time
distribution is also called survival probability and is obtained as
\begin{equation}
P_{surv}={1 \over N} \sum_{j=1}^N N_{rec}(n)~,
\label{eq13}
\end{equation}
where, the summation is taken along an ensemble of $N=10^6$ different
initial conditions. The term $N_{rec}(n)$ indicates the number of initial
conditions that do not escape through the hole at $V_{\rm hole}=100$ (i.e.
recur), until a collision $n$. The ensemble of initial conditions was set
for a constant velocity as $V_0=2\pi$ while $10^6$ phase were distributed 
evenly in $\phi_0\in[2.8,3.2]$.

\begin{table}[t]
{
\begin{tabular}{|c|c|c|c|c|c|} \hline \hline\hline
$\epsilon$&$\epsilon-\epsilon_c$&$-\zeta$&$-\gamma$ \\
\hline
$1.40$&$1.1557025$&$1.404(6)E-3$&$-$ \\
\hline
$1.30$&$1.057352$&$1.036(3)E-3$&$-$ \\
\hline
$1.20$&$0.957025$&$7.219(5)E-4$&$-$ \\
\hline
$1.10$&$0.857025$&$4.675(3)E-4$&$2.92(1)$ \\
\hline
$1.00$&$0.757025$&$3.430(7)E-4$&$2.18(1)$ \\
\hline
$0.90$&$0.657025$&$2.739(2)E-4$&$1.95(1)$ \\
\hline
$0.80$&$0.557025$&$2.105(2)E-4$&$1.625(9)$ \\
\hline
$0.70$&$0.457025$&$1.218(4)E-4$&$1.73(3)$ \\
\hline
$0.60$&$0.357025$&$5.260(1)E-5$&$2.16(2)$ \\
\hline
$0.575$&$0.332025$&$4.387(9)E-5$&$1.79(3)$ \\
\hline
$0.55$&$0.307025$&$3.71(7)E-5$&$1.52(1)$ \\
\hline
$0.525$&$0.282025$&$3.101(7)E-5$&$1.70(1)$ \\
\hline
$0.50$&$0.257025$&$2.463(5)E-5$&$1.91(9)$ \\
\hline
$0.475$&$0.232025$&$2.280(1)E-5$&$1.29(1)$ \\
\hline
$0.45$&$0.207025$&$1.408(3)E-5$&$1.71(1)$ \\
\hline
$0.425$&$0.182025$&$7.654(3)E-6$&$1.45(2)$ \\
\hline
$0.40$&$0.157025$&$5.73(9)E-6$&$1.90(5)$ \\
\hline
$0.375$&$0.132025$&$3.25(3)E-6$&$1.90(2)$ \\
\hline
$0.35$&$0.107025$&$1.536(4)E-6$&$1.84(1)$ \\ \hline
\hline
\hline
\end{tabular} }
\caption{Exponents obtained from numerical fitting for the curves of
$P_{surv}$ for different values of $\epsilon$.}
\label{Tab1}
\end{table}

It is known in the literature that if a system has fully chaotic behavior
the curves of $P_{surv}$ have an exponential decay \cite{ref40}. However,
when a mixed dynamics is observed in the phase space, the curves of
$P_{surv}$ may present different behaviors that may include: (i) a power
law decay \cite{ref41} or; (ii) a stretched exponential decay \cite{ref42}.
For the bouncer model which has a mixed phase space the curves of $P_{surv}$
may present either behaviors, depending on the parameter $\epsilon$ and the
set of initial conditions, as shown in Fig. \ref{fig6}.
We see a transition in the behavior of the curves of $P_{surv}$ as the
parameter $\epsilon$ is decreased. For large values of $\epsilon$ as for
example $\epsilon=1.4$ and $\epsilon=1.3$, the phase space has quite few
islands and the chaotic sea is dominant over the dynamics. It is therefore
expected an exponential decay in the curves of $P_{surv}$, as shown in Figs.
\ref{fig6}(a,b). As the parameter $\epsilon$ is getting smaller, more and
more stability islands appear in the phase space leading to the appearance
of more and more sticky regions. With these stable regions around in
the phase space, a change in the behavior of the curves of $P_{surv}$ is
expected. For values of $\epsilon<1$, we may observe a combination of decays
in the curves of $P_{surv}$. Firstly the  curves exhibit an exponential
decay and suddenly they change to a slow decay that we observed to be
described as a power law which marks the presence of orbits in stickiness
regime \cite{ref41}.

Considering the curves of the survival probability shown in Fig. \ref{fig6},
a numerical fitting can be made therefore according to: (i) the exponential
decay is given as $P_{surv}(n)\propto \exp(n\zeta)$ while; (ii) the power
law decay is described by $P_{surv}(n)\propto n^{\gamma}$ where $\zeta$ and
$\gamma$ are respectively the exponents for exponential and power law time
decays. Table \ref{Tab1} shows the set of exponents for different values of
the control parameter $\epsilon$.

\begin{figure}[t]
\includegraphics[width=1.0\linewidth]{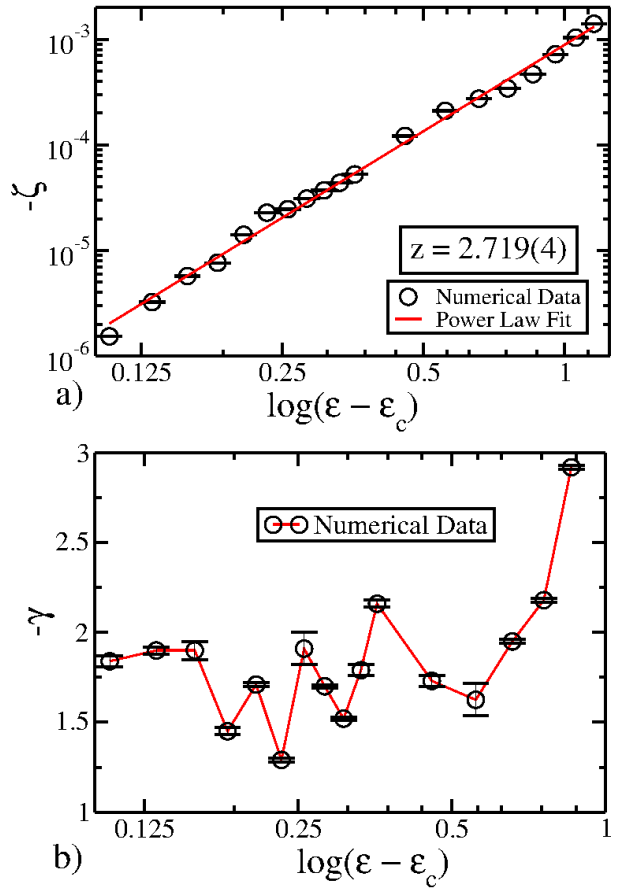}
\caption{(Color online). Plot of $-\zeta$ and $-\gamma$ as a function of
$\epsilon-\epsilon_c$.} 
\label{fig7}
\end{figure}

\begin{figure}[t]
\includegraphics[width=1.0\linewidth]{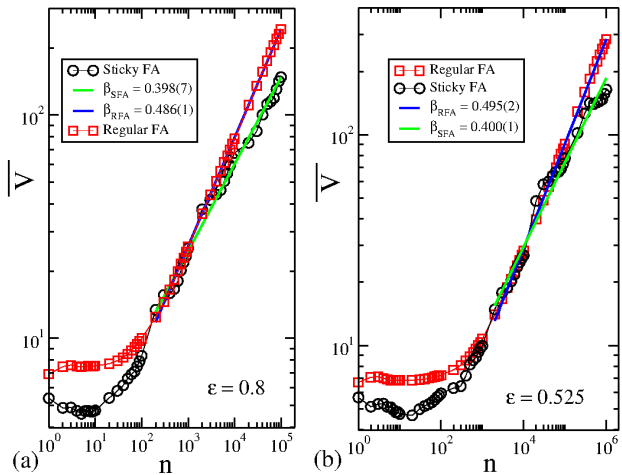}
\caption{(Color online). Plot of $\overline{V}$ as function of $n$ for: (a)
$\epsilon=0.8$ and (b)$\epsilon=0.525$. One can see two distinct growth
exponents for Regular Fermi Acceleration and Sticky Fermi Acceleration.
 Such difference can be undesrtood as sticky
orbits acting as a slowing mehanism for FA.} 
\label{fig8}
\end{figure}

We see that as the parameter $\epsilon$ decreases the exponential decay of
the curves of $P_{surv}$ also suffer a change. The exponent $\zeta$
decreases too as $\epsilon$ decreases, a result which is quite expected
given the periodic regions of the phase space are getting larger and larger.
Figure \ref{fig7}(a) shows the behavior of the exponent $\zeta$ as a
function of $(\epsilon-\epsilon_c)$. Looking at Fig. \ref{fig7}(a) we see
that the exponent $\zeta$ can be described by a power law of the type
$-\zeta\propto(\epsilon-\epsilon_c)^z$ and that the slope of the power law
is given by $z=2.719(4)$. 

The exponent $\gamma$ however does not show the mathematical beauty as
observed for the exponent $\zeta$. The slower decay observed in the curves
of the survival probability is indeed due to sticky regions present in the
phase space. For our simulations, most of the slower decay was characterized
as a power law. Indeed in the literature, it is known that the power law
decay, for such cumulative recurrence time distribution for other dynamical
systems \cite{ref43,ref44} which includes also billiards systems
\cite{ref41,ref45,ref46,ref47,ref48} is set in a range of $-\gamma \in [1.5,2.5]$ and that our
results match this range. We stress however that the total understanding and
this behavior is still an open problem and extensive theoretical and
numerical simulations, are required to describe its behavior properly.

Let us now address specifically the assumption that stickness may affect the
phenomenon of Fermi acceleration. Indeed the trapping dynamics of the
particles around stable regions makes the unlimited energy growth slower
than the usual. For a large set of initial conditions that lead the dynamics
of the particle to present diffusion in the velocity, the average velocity
$\bar{V}$ is described by $\bar{V}\propto \sqrt{n}$. However we expect the
initial conditions that spend large time trapped in sticky regions lead the
slope of growth to be smaller than $1/2$. This is indeed true and figure
\ref{fig8} confirms this assumption. The curves shown in bullets in both
Fig. \ref{fig8}(a,b) are named as Regular Fermi Acceleration (RFA) and were
obtained for evolution of the initial conditions which produce a fast decay
in the survival probability (those along the exponential decay in Fig.
\ref{fig6}) and as expected, an exponent of $\cong 0.5$ was obtained. On the
other hand, the curves plotted as squares show the evolution of initial
conditions chosen in the very final tail of the power law decay shown in
Fig. \ref{fig6} and are called as Sticky Fermi Acceleration (SFA). Power
law fitting furnish slopes $0.398(7)$ for (a) and $0.400(1)$ for (b).
These curves indeed give support for our claim that sticky regions slow
down the Fermi acceleration.

\section{Final Remarks and Conclusions}
\label{sec4}

The dynamics of the bouncer model was investigated by using a two
dimensional measure preserving mapping controlled by a single control
parameter $\epsilon$. For $\epsilon=0$ the system is integrable while it is
non integrable for $\epsilon\ne 0$. As soon as $\epsilon$ increases, the
periodic regions of the phase space reduce given rise to chaotic dynamics.
Indeed for $\epsilon>\epsilon_c$ invariant tori are not observed in the
phase space while periodic regions are still observed. The influence of
sticky regions also reduces with the increase of $\epsilon$. Our numerical
investigation of the FTLE spectrum distribution give support that trapping
dynamics is often observed in the phase space and is confirmed by the
secondary peaks of the FTLE distribution. The survival probability is
characterized by two decaying regimes: (1) for strong chaotic dynamics, the
decay is given by an exponential type while (2) it changes to a slower
decay marked by a power law type when mixed dynamics is present in the phase
space. Finally, 
according to the results shown in Fig.\ref{fig8}, we see that when a strong 
regime of stickiness is present in the system, it acts as a slowing 
mechanism for FA.  As with the survival probability, it would 
interesting to investigate whether the stickiness associated with mixed 
phase space in general models leads to a universal ``slowing 
exponen''.

\acknowledgments
ALPL acknowledges CNPq for financial support. CPD thanks Pr\'o Reitoria de
Pesquisa - PROPe/UNESP and DEMAC for hospitality during his stay in Brazil.
ILC thanks CNPq and FAPESP. EDL acknowledges CNPq, FAPESP and FUNDUNESP,
Brazilian agencies. This research was supported by resources supplied by the
Center for Scientific Computing (NCC/GridUNESP) of the S\~ao Paulo State
University (UNESP).

\end{document}